\documentclass[%
 amsmath,amssymb,
 longbibliography,
floatfix,notitlepage,12pt
]{revtex4-1}

\usepackage{graphicx}% Include figure files
\usepackage{dcolumn}% Align table columns on decimal point
\usepackage{bm}% bold math
\usepackage{mediabb}
\newcommand{\Iztot}{I_z^{\mbox{\scriptsize tot}}}

\begin{document}
%\preprint{APS/123-QED}

\title{Universal transport and resonant current from chiral magnetic effect}


\author{Hiroyuki Fujita and Masaki Oshikawa}
\affiliation{Institute for Solid State Physics,the  University of Tokyo, 5-1-5 Kashiwanoha, Kashiwa-shi, Chiba 277-8581, Japan}%Lines break automatically or can be forced with \\

%\begin{abstract}
%Chiral magnetic effect (CME) is a phenomenon in which an electric current proportional to the external magnetic field is predicted. It was originally derived for relativistic Weyl fermions in three dimensions but recent studies on Weyl semimetals renewed the interest on the subject, as a realistic problem which can be studied experimentally. Here we show that transport properties of Weyl semimetals supporting CME will be strongly affected by the consistency with the law of electromagnetism.   \end{abstract}

\maketitle

\textbf{
For relativistic Weyl fermions in $3+1$ dimensions,
an electric current proportional to the external magnetic
field is predicted.
This remarkable phenomenon is called 
Chiral Magnetic Effect (CME).
Recent studies on ``Weyl semimetals'' in condensed matter physics
renewed the interest on the subject
as a realistic problem which can be investigated experimentally.
Here we show that actual transports in Weyl semimetals
supporting CME cannot be discussed without proper consideration
of the law of electromagnetism.
That is, the current and electromagnetic fields are not only
related by the CME but also via electromagnetic induction
and Amp\`{e}re's law.
These intertwinings lead to observable transport properties
governed by CME rather different from what one would expect
naively from CME.
First, even in the absence of an external magnetic field,
CME leads to a material-independent, universal effective capacitance
under certain conditions.
Moreover, the induced current by a time-dependent external magnetic
field can be resonantly enhanced reflecting a
formation of electromagnetic standing waves. Our results imply that
electromagnetism plays an essential role in electromagnetic properties
of topological semimetals and its considerations is essential for the
applications of CME to future electronics.}

Interface between quantum field theory and condensed matter physics
has been a source of many important developments.
Weyl fermions and accompanying chiral anomaly
is a particularly notable example.
Their realization and consequences in condensed matter physics
were discussed earlier as an offspring of the
Nielsen-Ninomiya theorem in lattice field theory~\cite{Nielsen1983389},
and then were revisited later in the context of
quantum phase transitions~\cite{1367-2630-9-9-356}. 
More recently, a realization of Weyl fermions were predicted 
in pyrochlore iridates based on an \emph{ab initio}
calculation, and dubbed as \emph{Weyl semimetals}~\cite{PhysRevB.83.205101}.
This sparked a remarkable surge
in the theoretical and experimental activities.
Since then, Weyl semimetals have been reported to
be experimentally confirmed in TaAs, NbP, and
NbAs~\cite{PhysRevX.5.031013,
WeylsemimetalphaseinthecentrosymmetriccompoundTaAs,
ObservationofWeylnodesinTaAs, 2015arXiv150907465B, Xu:2015aa},
for example.

%Nowadays, materials which possess Weyl fermions are called
%topological semimetals\cite{1367-2630-9-9-356} which include Dirac semimetals\cite{PhysRevLett.108.140405} and Weyl semimetals\cite{PhysRevB.83.205101}. 
%CME can be relevant to both topological semimetals but CME in its original sense:``current only with a magnetic field" is possible only in Weyl semimetals.

These developments bring up the hope that CME~\cite{PhysRevD.78.074033}
may be also observed experimentally~\cite{0953-8984-27-11-113201,PhysRevD.86.045001,PhysRevB.89.075124,PhysRevB.89.081407,Pallab_noncentro,
PhysRevB.92.161110, PhysRevB.91.115203, Kharzeev2014133}.
Indeed, although it is still
controversial~\cite{PhysRevB.92.075205, 2014arXiv1412.5168K}, the giant
negative quadratic magnetoresistance observed in Dirac
semimetals~\cite{Xiong23102015, PhysRevX.5.031023} strongly suggests that
the physics of relativistic fermions, in particular the chiral
anomaly~\cite{PhysRevB.88.104412}, is indeed taking place in these materials.
However, at
present there is no direct observation of CME in the original sense
of an electric current driven by the external magnetic field.
In the following, we explore possible observations of CME
in terms of the induced current.
In lattice models CME vanishes as a static effect in
the thermal equilibrium~\cite{PhysRevLett.111.027201, PhysRevB.88.125105}
but may exist
in the non-equilibrium limit~\cite{Pallab_noncentro,
PhysRevB.91.115203}. Thus we consider the CME at a finite frequency
$\omega$, characterized by a finite chiral magnetic
conductivity $\sigma_{ch}(\omega)$ so that
\begin{equation}
\bm{j}_{ch}(\omega) = \sigma_{ch}(\omega)\bm{B}(\omega) .
\label{eq.CME}
\end{equation}
However, there is an important issue to be addressed carefully
for any discussion of possible observation of the CME:
the electric and magnetic fields, and electric charge and current densities
have to obey the laws of electrodynamics.

CME can be regarded as a
Chern-Simons type extension of electromagnetism.
As is known~\cite{PhysRevLett.58.1799, PhysRevLett.108.161803} in such cases
the law of electromagnetism may lead to nontrivial responses. In this
letter, by solving the Maxwell-Chern-Simons (MCS) equations we
demonstrate that CME will qualitatively change transport properties of
matter, in a rather unexpected way.
As we will show, the physically observed conductance is
not simply given by the chiral magnetic conductivity
$\sigma_{ch}(\omega)$, even when it is governed by the CME.
Our results imply that the electromagnetism is fundamental for
actual transports in Weyl semimetals and that proposals for their
applications to future electronics~\cite{PhysRevB.88.115119} need
careful considerations on this issue.

\begin{figure}[htbp]
\centering
   \includegraphics[width = 60mm]{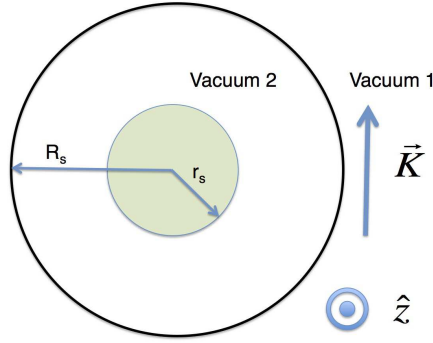}
   \caption{Schematic picture of our setup. A solenoid (radius $R_{s}$) is represented as a surface current between the two vacua. Inside them, a cylindrical sample (radius $r_{s}$) is placed.}
      \label{solenoid figure}
\end{figure}
We consider an infinitely long cylinder of the sample with nonzero chiral
magnetic conductivity, placed inside an infinitely long solenoid (FIG.1).
While any realistic system is of a finite length, 
we expect that our analysis essentially applies to such systems,
as long as there is a non-vanishing uniform component along the
cylinder axis in all the physical quantities including the current.

Fourier transforming in time, MCS
equations inside the sample takes the following form:
\begin{align}
\nabla \cdot \bm{B}&=0 \label{maxwell 1}\\ \nabla \cdot \bm{E} &=
\frac{\rho}{\epsilon_{}} \label{maxwell 2} \\ \nabla \times \bm{E} -i
\omega \bm{B}&= 0 \label{maxwell 3} \\ \nabla \times \bm{B}+i \omega
\epsilon \mu \bm{E} &= \mu \sigma \bm{E}+ \mu \sigma_{ch} \bm{B}
\label{maxwell 4}
\end{align}
where $\rho$ is charge density, $\omega$ is frequency,
$\sigma_{ch}(\omega)$ and $\sigma(\omega)$ respectively
are chiral magnetic and ordinary longitudinal conductivity.
For simplicity we do not consider
the Hall effect. In the following, we assume $\mu \sigma \gg
\omega/c^{2}$ and drop $i \omega/c^{2} \bm{E}$ inside the sample.
To solve the MCS equations, we introduce the following dimensionless
constant
\begin{align}
\delta= \frac{\omega \sigma}{\mu \sigma_{ch}^{2}}
\end{align}
and consider a perturbative expansion of fields in terms of $\delta$.
According an estimation for a lattice model\cite{Pallab_noncentro}, when
the energy difference between Weyl nodes is $\Delta$ [meV],
$\sigma_{ch}$ is of the order of $10^{7}$ $\Delta$
$[\mathrm{A}/\mathrm{m^{2} T}]$.
Since
semimetals typically have $\sigma$ $\sim$ $10^{6}$
$[S/\mathrm{m}]$,
$|\delta|$ is about $\delta \sim 10^{-2}\omega/\Delta^{2}$ 
for the frequency $\omega$ [Hz],
if we assume $\mu \simeq \mu_{0}$.
Therefore, if we assume $\Delta \sim 100$ [meV], our
analysis works up to about $\omega \sim 100$ [kHz].

The geometry
of our setup leads to the following conditions on fields,
in terms of the cylindrical coordinates $(r,\phi,z)$:
\begin{align}
\frac{\partial}{\partial\phi} \bm{E}=
\frac{\partial}{\partial\phi}\bm{B}=\frac{\partial}{\partial z}\bm{E}=
\frac{\partial}{\partial z}\bm{B} = B_r=E_r=0.
\label{symmetrycondition}
\end{align}
Under these conditions, the equations (\ref{maxwell 1}) and
(\ref{maxwell 2}) are trivial. We shall solve
the remaining equations \eqref{maxwell 3} and \eqref{maxwell 4}
in the small-$\delta$ expansion.
First we define a length scale
$ l_{\mathrm{CML}} =\pi/(\mu \sigma_{ch})$,
which we call {\it chiral magnetic length} (CML).  
Here the factor $\pi$ is introduced for a later convenience.
Scaling the radial coordinate $r$ as
$\displaystyle x = \pi r / l_{\mathrm{CML}}$,
at $\mathrm{O}(\delta^{0})$ we obtain
\begin{align}
B_{z}(x) &= a J_{0}(x)\label{EMfields 1}\\
B_{\phi}(x) &= a J_{1}(x)
\label{EMfields 2}\\
\bm{E}(x) &= \frac{i\omega}{\mu \sigma_{ch}}
\bm{B}(x),
\label{EMfields 3}
\end{align}
where $J_{i}(x)$ represents $i$-th Bessel function
and $a$ is an undetermined constant.
Given the solutions \eqref{EMfields 1} and \eqref{EMfields 3},
the parameter $\delta$ can be expressed as
\begin{align}
\delta \simeq 
\left|
\frac{\sigma E_{z}}{\sigma_{ch} B_{z}}
\right|
=
\left|
\frac{i \omega \sigma}{\mu  \sigma_{ch}^{2}}
\right|,
\label{delta_phys}
\end{align}
namely by
the ratio of the ordinary current and the current due to CME.
Therefore our analysis is physically a large CME expansion.

% The unknown coefficient $a$ is determined by solving electromagnetic
% boundary conditions:
% \begin{align}
% (E_{\phi}^{\mathrm{cyl}} - E_{\phi}^{\mathrm{vac 2}})|_{r = r_{s}} = 0
% \nonumber \\ (E_{z}^{\mathrm{cyl}} - E_{z}^{\mathrm{vac
% 2}})|_{r = r_{s}} =0, \label{bc}
% \end{align}
% where``cyl'' and ``vac 2'' denote fields inside and outside the sample
% and $r_{s}$ is the radius of the cylinder.
Electromagnetic fields in the
``vacuum 2'' region between the sample and the solenoid are derived in
the Supplementary Material.  In the lowest order of $\omega r_{s}/c$ and
$\omega R_{s}/c$, they are given by:
\begin{align}
B_{z}^{\mathrm{vac 2}}(r) &= -\frac{2 i}{\pi c}\log \left[
\frac{\omega}{c}r \right]e_{2}^{Y} + K \label{Bz expanded}\\
B_{\phi}^{\mathrm{vac 2}}(r) &= -\frac{2 c}{\pi
\omega}\frac{1}{r}b_{2}^{Y} \label{Bphi expanded}\\
E_{z}^{\mathrm{vac 2}}(r) &= \frac{2 i c}{\pi}\log \left[
\frac{\omega}{c} r \right] b_{2}^{Y}.  \label{Ez expanded}\\
E_{\phi}^{\mathrm{vac 2}}(r) &=-\frac{2c }{\pi
\omega}\frac{1}{r}e_{2}^{Y} +\frac{i \omega }{2} K r,
\label{Ephi expanded}
\end{align}
 where $e_{2}^{Y}$ and $b_{2}^{Y}$ are undetermined constants. $K/\mu_{0}$ is
   the current flowing the solenoid in the angular direction per length
of the cylinder.
The three unknown coefficients, $a$, $e_{2}^{Y}$, and $b_{2}^{Y}$, are
related by the two boundary conditions
\begin{equation}
 (E_{\phi}^{\mathrm{cyl}} - E_{\phi}^{\mathrm{vac 2}})|_{r = r_{s}} =
(E_{z}^{\mathrm{cyl}} - E_{z}^{\mathrm{vac 2}})|_{r = r_{s}} =0,
\label{bc}
\end{equation}
where 
$r_s$ is the radius of the cylindrical sample, and 
``cyl'' denotes the fields inside the sample.
Thus there is one remaining undetermined constant.

Below, we discuss transport properties of the cylinder based on the
solutions we obtained. The total current $\Iztot$
flowing along the cylindrical axis is simply given by
\begin{align}
 \Iztot = -\frac{4 c}{\mu_{0}\omega}b_{2}^{Y},
\label{relation j-b}
\end{align}
thanks to Amp\`{e}re's law and Eq.~\eqref{Bphi expanded}.
% This is easily derived by integrating (\ref{maxwell 4}) taking into
% account a surface current coming from the discontinuity of $B_{\phi}$ at
% the surface of the cylinder: $j_{z}^{\mathrm{surface}} = 2\pi
% r_{s}(B_{\phi}^{\mathrm{vac 2}}/\mu_{0} - B_{\phi}^{cyl}/\mu) $. Notice
% that the bulk current is cancelled with the second term of the surface
% current and relation (\ref{relation j-b}) is independent of the magnetic
% susceptibility of the cylinder $\mu$.

First, we consider a situation
where a definite amount of current $\Iztot$ is injected to the
cylinder and the voltage drop along the cylindrical axis is
measured. We can then define the complex electric conductance
(admittance) $G$ by the ratio of $\Iztot$ and $E_{z}L$,
in which the residual undetermined constant $b_{2}^{Y}$ cancels.
%In the lowest order of
%$\mu \sigma_{ch} r_{s}$, 
The admittance $G$
of a cylinder with length $L$ is then uniquely determined
as $G = - i \omega C$, where
\begin{align}
 C &= \frac{i I_{z}^{tot}}{\omega L E_{z}}
 \simeq \frac{2\pi}{ \mu_{0} \omega^2 L \,
 \log \left[ \frac{c}{\omega r_s} \right]},
 \label{univ capacitance}
 \end{align} 
is real and positive in a realistic setting $\omega r_s \ll c$.
That is, with the current leading the voltage by the phase $\pi/2$,
the response in this order is purely
capacitive with the effective capacitance~\eqref{univ capacitance}.
Qualitatively, this effective capacitive behavior may be understood
as a consequence of the proportionality~\eqref{EMfields 3} between
the magnetic and electric fields in the sample, and the local CME
relation~\eqref{eq.CME}.
The proportionality~\eqref{EMfields 3} implies that
the admittance $G$, being the ratio between
the voltage drop along the cylinder axis and the CME-induced current,
does not depend on the solenoid current $K$.
Thus, in terms of the admittance,
the same capacitive transport governed by CME is
observed even without applying any magnetic field externally ($K=0$)!
We note that, while the setting and actual responses are different,
there is a similarity between the capacitive transport and
the screening of the electric field in ``axionic''
materials which is also related
to the MCS~\cite{PhysRevLett.108.161803}.

On the other hand, the quantitative estimate of the effective
capacitance requires the analysis of the MCS equation with proper
boundary conditions as described above.
It is remarkable that the final result~\eqref{univ capacitance} does not
contain any of the material parameters: $\epsilon$, $\mu$, $\sigma$, and
$\sigma_{ch}$.  Since our analysis is based on the macroscopic
electromagnetism, all the details of the system are renormalized in
those parameters.  Therefore, the effective
capacitance~(\ref{univ capacitance})
is a material independent, universal quantity.
It should be stressed that, although the formula~\eqref{univ capacitance}
does not  depend explicitly on the chiral magnetic conductivity $\sigma_{ch}$,
 it \emph{is} a
 consequence of CME as it is obtained in the leading order in the large
 CME expansion.
%  We note that Eq.~\eqref{univ capacitance}
%  also does not depend on $K$ so we can take $K = 0$. It means that,
%  even in the absence of the external magnetic field, the AC transport
%  properties of the system may be governed by CME.
% At the same time, the
 In any case, the
 result is quite different from what would be expected naively from CME
 without taking the law of electromagnetism into account. 

Next, we evaluate the response of the total current $\Iztot$
to the applied external magnetic field induced by the solenoid current $K$.
Without taking the electromagnetism into account, it would be
simply given by the cross section of the cylinder times the current
density: $I_{z}^{tot} = \pi r_{s}^{2} \sigma_{ch} K$. However, we shall
see that the result is rather different.
In this case we have to specify the undetermined constant explicitly by
imposing an additional condition. Here we assume that the
``back-reaction'' terms proportional to $e_{2}^{Y}$, which can only
exist in the presence of the sample, are small compared to terms
proportional to $K$. Then we have:
\begin{align}
 I_{z}^{tot} &= \frac{ \pi r_{s} J_{0}(\mu \sigma_{ch} r_{s})}{\mu_{0}
  J_{1}(\mu \sigma_{ch} r_{s})\log \left[ \frac{c}{\omega r_s} \right]} K.
 \label{resonance of current}
\end{align}
This current vanishes for $\omega \rightarrow 0$ due to the
divergence of the denominator. As is known, CME in lattice models does
not exist in thermal equilibrium. Here we obtained a stronger result:
even if $\sigma_{ch}(\omega \rightarrow 0)$ is non-vanishing, the total
current induced by a solenoid vanish (logarithmically)
for $\omega \rightarrow 0$ if we
take into account the law of electromagnetism.

In FIG. 2, we show
$I_{z}^{tot}$ for the following parameters: $\mu \sigma_{ch} =1
\mathrm{\, [mm^{-1}}], \, \omega = 100 \mathrm{\,[Hz]}, \, B = 1
\mathrm{\,[Gauss]}$.
\begin{figure}[htbp]
\centering \includegraphics{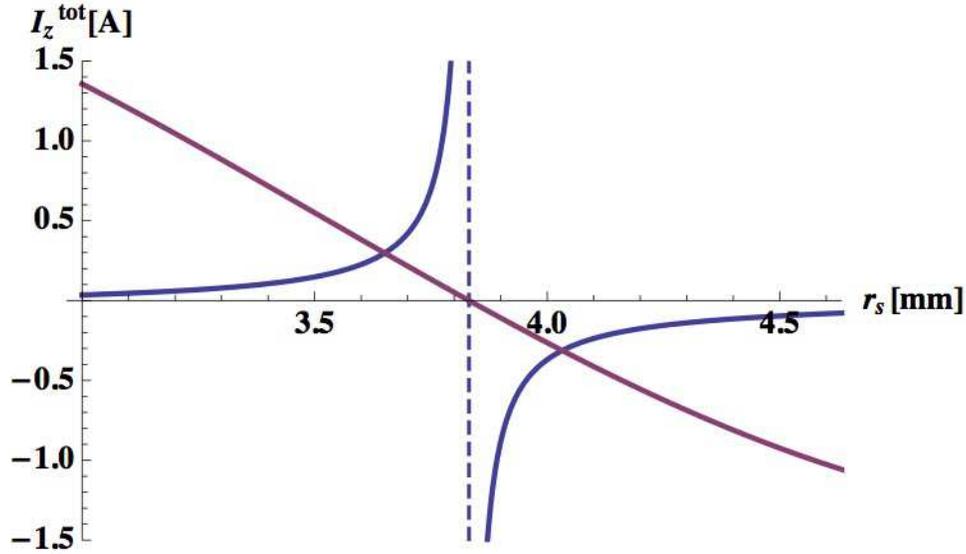}
   \caption{Blue line: Total current $I_{z}^{tot}$[A] for parameters
   $\mu \sigma_{ch} =1 \mathrm{\, [mm^{-1}}], \omega = 100
   \mathrm{\,[Hz]}, B = 1 \mathrm{\,[Gauss]}$. Red: $J_{1}(\mu
   \sigma_{ch} r_{s})$ in an arbitrary unit. The current is resonantly
   enhanced for $r_{s}$ satisfying $J_{1}(\mu \sigma_{ch} r_{s}) =0$
   represented by a dashed line.}
\end{figure}
 It is clear that the current is resonantly enhanced if $J_{1}(\mu
 \sigma_{ch}r_{s}) = 0$ is satisfied. Using the chiral magnetic length
 $l_{\mathrm{CML}}$, the condition is represented as:
\begin{align}
\frac{r_{s} }{l_{\mathrm{CML}}}\simeq(\mathbb{Z}+ \frac{1}{4})
\label{resonance}
\end{align}
since $J_{1}(y) \rightarrow \sqrt{2/\pi y}\cos(y - 3\pi/4 )$ for $y
\rightarrow \infty$. $l_{\mathrm{CML}}$ is the characteristic length
scale of electromagnetic fields inside the sample so the resonances
correspond to formation of two-dimensional electromagnetic standing
waves. As we noted, $\sigma_{ch}$ is about $10^{7}$ $\Delta$ where
$\Delta$ is the energy difference between Weyl nodes in the unit of
meV. $l_{\mathrm{CML}} = \pi/(\mu \sigma_{ch})$ is then given by
$\sim100 \pi / \Delta$ [mm]. Therefore, when $\Delta =
\mathrm{100}$[meV], the resonances will be achieved for $r_{s} \simeq
\pi(\mathbb{Z} + 1/4)$[mm]. \par

%Without $\sigma_{ch}$, the length scale of electromagnetic fields is $1/\sqrt{\mu \sigma \omega}$ and this is much larger than $l_{\mathrm{CML}}$ so even if such a standing wave is possible in principle, practically its quite hard to achieve in ordinary materials.\par
Even away from the resonances, our result is still peculiar. If we
assume $\mu \sigma_{ch} r_{s} \ll 1$, in the lowest order we have:
\begin{align}
I_{z}^{tot}\simeq -\frac{2 l_{\mathrm{CML} }}{\mu_{0} \log \left[
\frac{\omega}{c} r_{s} \right]}K.
\end{align}
This means that in this parameter region, the induced current is
monotonically ``decreasing'' as a function of $\sigma_{ch}$.
\begin{acknowledgments}
We thank Satoru Nakatsuji for stimulating discussions which led
to the present project.
Useful discussions with Pallab Goswami and Takahiro Tomita
are also gratefully acknowledged.
This work has been partially supported by
JSPS Grant-in-Aid for Scientific Research (KAKENHI)
No. 15H02113 and JSPS Strategic International Networks Program
No. R2604 ``TopoNet''. HF was supported by Japan Society for the Promotion of Science through Program for Leading Graduate Schools (ALPS).
A part of this work was carried out at Kavli Institute for Theoretical
Physics, University of California, Santa Barbara, supported by the
US National Science Foundation under Grant No. NSF PHY11-25915.
\end{acknowledgments}

\newpage
\appendix

\section{Electromagnetic fields in the vacuum}
To implement a solenoid in the framework of the Maxwell's equations, we
divide the vacuum region outside the material into
two vacua (FIG. 1), inside and outside the solenoid.
We then impose the following boundary conditions between them:
\begin{align}
E_{z}^{\mathrm{vac} 2}(r = R_{s}) - E_{z}^{\mathrm{vac}
1}(r=R_{s}) &=0\label{sol bc1} \\ E_{\phi}^{\mathrm{vac} 2}(r =
R_{s}) - E_{\phi}^{\mathrm{vac} 1}(r=R_{s}) &= 0 \label{sol bc2}\\
B_{z}^{\mathrm{vac} 2}(r = R_{s}) - B_{z}^{\mathrm{vac}
1}(r=R_{s}) &= K \label{sol bc3}\\ B_{\phi}^{\mathrm{vac} 2}(r =
R_{s}) - B_{\phi}^{\mathrm{vac} 1}(r=R_{s}) &= 0 \label{sol bc4}
\end{align}
where ``vac 1'' and ``vac 2'' label the two vacua, $K$ models the current
flowing the solenoid in the angular direction, and $R_{s}$ is the radius
of the solenoid.  Using (\ref{symmetrycondition}), the fields are
determined as a solution of the Maxwell's equations in the vacuum:
\begin{align}
E_{z}^{vac 1, 2}(r) &= i c \left[\,b_{1, 2}^{J} J_{0}\left(\frac{\omega}{c}r \right) +\, b_{1, 2}^{Y} Y_{0}\left(\frac{\omega}{c}r \right) \right] \\
B_{\phi}^{vac 1, 2}(r) &= b_{1, 2}^{J} J_{1}\left(\frac{\omega}{c}r \right) + b_{1, 2}^{Y} Y_{1}\left(\frac{\omega}{c}r \right) \\
B_{z}^{vac 1, 2} (r)&=  \frac{1}{i c}\left[ e_{1, 2}^{J} J_{0}\left(\frac{\omega}{c}r \right) + e_{1, 2}^{Y} Y_{0}\left(\frac{\omega}{c}r \right) \right] \\
E_{\phi}^{vac 1, 2}(r) &= e_{1, 2}^{J} J_{1}\left(\frac{\omega}{c}r \right) + e_{1, 2}^{Y} Y_{1}\left(\frac{\omega}{c}r \right).
\end{align}
Here $e_{1, 2}^{J,Y}$ and $b_{1, 2}^{J,Y}$ are undetermined constants
and $J_{i}, Y_{i}$ are Bessel functions of the first and second kind
respectively. To determine the unknown coefficients, first we require
causality in ``vacuum 1'' region. In other words, we require the
non-existence of ``incoming waves'' from the infintiy. This is equivalent
to force the fields to be regular in $\mathrm{Im}\, \omega > 0$. Because
both $J_{i}\left(\frac{\omega}{c}r \right)$ and
$Y_{i}\left(\frac{\omega}{c}r \right)$ diverge for $\omega
\rightarrow i \infty$, to meet the requirement, we have to take a linear
combination of them to cancel out the divergence (notice that because we
can take we can take $r \rightarrow \infty$ in this region, the
unknown coefficients cannot regularize the divergence). Then it turns
out that the Hankel function: $H_{i}\left(\frac{\omega}{c}r \right) =
J_{i}\left(\frac{\omega}{c}r \right) + i
Y_{i}\left(\frac{\omega}{c}r \right)$ is the only combination regular
in the upper half complex plane of $\omega$. Therefore, causality
requires $e_{1}^{Y} = i e_{1}^{J}$ and $b_{1}^{Y} = i b_{1}^{J} $.

Then (\ref{sol bc1})$\sim$(\ref{sol bc4}) give the following results with only two unknown coefficients $e_{2}^{Y}$ and $b_{2}^{Y}$:
 \begin{align}
E_{z}^{\mathrm{vac 2}} &= c \,b_{2}^{Y}H_{0}^{r} \\
B_{\phi}^{\mathrm{vac 2}} &= -i \, b_{2}^{Y} H_{1}^{r}\\
B_{z}^{\mathrm{vac 2}} &= - \frac{1}{c}e_{2}^{Y}H_{0}^{r}- \frac{\pi}{2 i c}\omega R_{s} H_{1}^{R_{s}} K J_{0}^{r} \nonumber \\
E_{\phi}^{\mathrm{vac 2}} &= - i e_{2}^{Y} H_{1}^{r}-\frac{\pi}{2}\omega R_{s} H_{1}^{R_{s}}K J_{1}^{r} \nonumber
 \end{align}
 where we have used abbreviated notations like $H_{i}(\omega R_{s}/c) \equiv H_{i}^{R_{s}}$ and $J_{i}(\omega R_{c}/c) \equiv J_{i}^{R_{s}}$.
\newpage
%\begin{thebibliography}

%\end{thebibliography}
\bibliography{Maxwell}% Produces the bibliography via BibTeX.

\begin{thebibliography}{27}%
\makeatletter
\providecommand \@ifxundefined [1]{%
 \@ifx{#1\undefined}
}%
\providecommand \@ifnum [1]{%
 \ifnum #1\expandafter \@firstoftwo
 \else \expandafter \@secondoftwo
 \fi
}%
\providecommand \@ifx [1]{%
 \ifx #1\expandafter \@firstoftwo
 \else \expandafter \@secondoftwo
 \fi
}%
\providecommand \natexlab [1]{#1}%
\providecommand \enquote  [1]{``#1''}%
\providecommand \bibnamefont  [1]{#1}%
\providecommand \bibfnamefont [1]{#1}%
\providecommand \citenamefont [1]{#1}%
\providecommand \href@noop [0]{\@secondoftwo}%
\providecommand \href [0]{\begingroup \@sanitize@url \@href}%
\providecommand \@href[1]{\@@startlink{#1}\@@href}%
\providecommand \@@href[1]{\endgroup#1\@@endlink}%
\providecommand \@sanitize@url [0]{\catcode `\\12\catcode `\$12\catcode
  `\&12\catcode `\#12\catcode `\^12\catcode `\_12\catcode `\%12\relax}%
\providecommand \@@startlink[1]{}%
\providecommand \@@endlink[0]{}%
\providecommand \url  [0]{\begingroup\@sanitize@url \@url }%
\providecommand \@url [1]{\endgroup\@href {#1}{\urlprefix }}%
\providecommand \urlprefix  [0]{URL }%
\providecommand \Eprint [0]{\href }%
\providecommand \doibase [0]{http://dx.doi.org/}%
\providecommand \selectlanguage [0]{\@gobble}%
\providecommand \bibinfo  [0]{\@secondoftwo}%
\providecommand \bibfield  [0]{\@secondoftwo}%
\providecommand \translation [1]{[#1]}%
\providecommand \BibitemOpen [0]{}%
\providecommand \bibitemStop [0]{}%
\providecommand \bibitemNoStop [0]{.\EOS\space}%
\providecommand \EOS [0]{\spacefactor3000\relax}%
\providecommand \BibitemShut  [1]{\csname bibitem#1\endcsname}%
\let\auto@bib@innerbib\@empty
%</preamble>
\bibitem [{\citenamefont {Nielsen}\ and\ \citenamefont
  {Ninomiya}(1983)}]{Nielsen1983389}%
  \BibitemOpen
  \bibfield  {author} {\bibinfo {author} {\bibfnamefont {H.B.}\ \bibnamefont
  {Nielsen}}\ and\ \bibinfo {author} {\bibfnamefont {Masao}\ \bibnamefont
  {Ninomiya}},\ }\bibfield  {title} {\enquote {\bibinfo {title} {{The
  Adler-Bell-Jackiw anomaly and Weyl fermions in a crystal}},}\ }\href
  {\doibase http://dx.doi.org/10.1016/0370-2693(83)91529-0} {\bibfield
  {journal} {\bibinfo  {journal} {Physics Letters B}\ }\textbf {\bibinfo
  {volume} {130}},\ \bibinfo {pages} {389 -- 396} (\bibinfo {year}
  {1983})}\BibitemShut {NoStop}%
\bibitem [{\citenamefont {Murakami}(2007)}]{1367-2630-9-9-356}%
  \BibitemOpen
  \bibfield  {author} {\bibinfo {author} {\bibfnamefont {Shuichi}\ \bibnamefont
  {Murakami}},\ }\bibfield  {title} {\enquote {\bibinfo {title} {{Phase
  transition between the quantum spin Hall and insulator phases in 3D:
  emergence of a topological gapless phase}},}\ }\href
  {http://stacks.iop.org/1367-2630/9/i=9/a=356} {\bibfield  {journal} {\bibinfo
   {journal} {New Journal of Physics}\ }\textbf {\bibinfo {volume} {9}},\
  \bibinfo {pages} {356} (\bibinfo {year} {2007})}\BibitemShut {NoStop}%
\bibitem [{\citenamefont {Wan}\ \emph {et~al.}(2011)\citenamefont {Wan},
  \citenamefont {Turner}, \citenamefont {Vishwanath},\ and\ \citenamefont
  {Savrasov}}]{PhysRevB.83.205101}%
  \BibitemOpen
  \bibfield  {author} {\bibinfo {author} {\bibfnamefont {Xiangang}\
  \bibnamefont {Wan}}, \bibinfo {author} {\bibfnamefont {Ari~M.}\ \bibnamefont
  {Turner}}, \bibinfo {author} {\bibfnamefont {Ashvin}\ \bibnamefont
  {Vishwanath}}, \ and\ \bibinfo {author} {\bibfnamefont {Sergey~Y.}\
  \bibnamefont {Savrasov}},\ }\bibfield  {title} {\enquote {\bibinfo {title}
  {{Topological semimetal and Fermi-arc surface states in the electronic
  structure of pyrochlore iridates}},}\ }\href {\doibase
  10.1103/PhysRevB.83.205101} {\bibfield  {journal} {\bibinfo  {journal} {Phys.
  Rev. B}\ }\textbf {\bibinfo {volume} {83}},\ \bibinfo {pages} {205101}
  (\bibinfo {year} {2011})}\BibitemShut {NoStop}%
\bibitem [{\citenamefont {Lv}\ \emph {et~al.}(2015{\natexlab{a}})\citenamefont
  {Lv}, \citenamefont {Weng}, \citenamefont {Fu}, \citenamefont {Wang},
  \citenamefont {Miao}, \citenamefont {Ma}, \citenamefont {Richard},
  \citenamefont {Huang}, \citenamefont {Zhao}, \citenamefont {Chen},
  \citenamefont {Fang}, \citenamefont {Dai}, \citenamefont {Qian},\ and\
  \citenamefont {Ding}}]{PhysRevX.5.031013}%
  \BibitemOpen
  \bibfield  {author} {\bibinfo {author} {\bibfnamefont {B.~Q.}\ \bibnamefont
  {Lv}}, \bibinfo {author} {\bibfnamefont {H.~M.}\ \bibnamefont {Weng}},
  \bibinfo {author} {\bibfnamefont {B.~B.}\ \bibnamefont {Fu}}, \bibinfo
  {author} {\bibfnamefont {X.~P.}\ \bibnamefont {Wang}}, \bibinfo {author}
  {\bibfnamefont {H.}~\bibnamefont {Miao}}, \bibinfo {author} {\bibfnamefont
  {J.}~\bibnamefont {Ma}}, \bibinfo {author} {\bibfnamefont {P.}~\bibnamefont
  {Richard}}, \bibinfo {author} {\bibfnamefont {X.~C.}\ \bibnamefont {Huang}},
  \bibinfo {author} {\bibfnamefont {L.~X.}\ \bibnamefont {Zhao}}, \bibinfo
  {author} {\bibfnamefont {G.~F.}\ \bibnamefont {Chen}}, \bibinfo {author}
  {\bibfnamefont {Z.}~\bibnamefont {Fang}}, \bibinfo {author} {\bibfnamefont
  {X.}~\bibnamefont {Dai}}, \bibinfo {author} {\bibfnamefont {T.}~\bibnamefont
  {Qian}}, \ and\ \bibinfo {author} {\bibfnamefont {H.}~\bibnamefont {Ding}},\
  }\bibfield  {title} {\enquote {\bibinfo {title} {{Experimental Discovery of
  Weyl Semimetal TaAs}},}\ }\href {\doibase 10.1103/PhysRevX.5.031013}
  {\bibfield  {journal} {\bibinfo  {journal} {Phys. Rev. X}\ }\textbf {\bibinfo
  {volume} {5}},\ \bibinfo {pages} {031013} (\bibinfo {year}
  {2015}{\natexlab{a}})}\BibitemShut {NoStop}%
\bibitem [{\citenamefont {Yang}\ \emph {et~al.}(2015)\citenamefont {Yang},
  \citenamefont {Liu}, \citenamefont {Sun}, \citenamefont {Peng}, \citenamefont
  {Yang}, \citenamefont {Zhang}, \citenamefont {Zhou}, \citenamefont {Zhang},
  \citenamefont {Guo}, \citenamefont {Rahn}, \citenamefont {Prabhakaran},
  \citenamefont {Hussain}, \citenamefont {Mo}, \citenamefont {Felser},
  \citenamefont {Yan},\ and\ \citenamefont
  {Chen}}]{WeylsemimetalphaseinthecentrosymmetriccompoundTaAs}%
  \BibitemOpen
  \bibfield  {author} {\bibinfo {author} {\bibfnamefont {L.~X.}\ \bibnamefont
  {Yang}}, \bibinfo {author} {\bibfnamefont {Z.~K.}\ \bibnamefont {Liu}},
  \bibinfo {author} {\bibfnamefont {Y.}~\bibnamefont {Sun}}, \bibinfo {author}
  {\bibfnamefont {H.}~\bibnamefont {Peng}}, \bibinfo {author} {\bibfnamefont
  {H.~F.}\ \bibnamefont {Yang}}, \bibinfo {author} {\bibfnamefont
  {T.}~\bibnamefont {Zhang}}, \bibinfo {author} {\bibfnamefont
  {B.}~\bibnamefont {Zhou}}, \bibinfo {author} {\bibfnamefont {Y.}~\bibnamefont
  {Zhang}}, \bibinfo {author} {\bibfnamefont {Y.~F.}\ \bibnamefont {Guo}},
  \bibinfo {author} {\bibfnamefont {M.}~\bibnamefont {Rahn}}, \bibinfo {author}
  {\bibfnamefont {D.}~\bibnamefont {Prabhakaran}}, \bibinfo {author}
  {\bibfnamefont {Z.}~\bibnamefont {Hussain}}, \bibinfo {author} {\bibfnamefont
  {S.~K.}\ \bibnamefont {Mo}}, \bibinfo {author} {\bibfnamefont
  {C.}~\bibnamefont {Felser}}, \bibinfo {author} {\bibfnamefont
  {B.}~\bibnamefont {Yan}}, \ and\ \bibinfo {author} {\bibfnamefont {Y.~L.}\
  \bibnamefont {Chen}},\ }\bibfield  {title} {\enquote {\bibinfo {title} {{Weyl
  semimetal phase in the non-centrosymmetric compound TaAs}},}\ }\href
  {http://dx.doi.org/10.1038/nphys3425} {\bibfield  {journal} {\bibinfo
  {journal} {Nat Phys}\ }\textbf {\bibinfo {volume} {11}},\ \bibinfo {pages}
  {728--732} (\bibinfo {year} {2015})}\BibitemShut {NoStop}%
\bibitem [{\citenamefont {Lv}\ \emph {et~al.}(2015{\natexlab{b}})\citenamefont
  {Lv}, \citenamefont {Xu}, \citenamefont {Weng}, \citenamefont {Ma},
  \citenamefont {Richard}, \citenamefont {Huang}, \citenamefont {Zhao},
  \citenamefont {Chen}, \citenamefont {Matt}, \citenamefont {Bisti},
  \citenamefont {Strocov}, \citenamefont {Mesot}, \citenamefont {Fang},
  \citenamefont {Dai}, \citenamefont {Qian}, \citenamefont {Shi},\ and\
  \citenamefont {Ding}}]{ObservationofWeylnodesinTaAs}%
  \BibitemOpen
  \bibfield  {author} {\bibinfo {author} {\bibfnamefont {B.~Q.}\ \bibnamefont
  {Lv}}, \bibinfo {author} {\bibfnamefont {N.}~\bibnamefont {Xu}}, \bibinfo
  {author} {\bibfnamefont {H.~M.}\ \bibnamefont {Weng}}, \bibinfo {author}
  {\bibfnamefont {J.~Z.}\ \bibnamefont {Ma}}, \bibinfo {author} {\bibfnamefont
  {P.}~\bibnamefont {Richard}}, \bibinfo {author} {\bibfnamefont {X.~C.}\
  \bibnamefont {Huang}}, \bibinfo {author} {\bibfnamefont {L.~X.}\ \bibnamefont
  {Zhao}}, \bibinfo {author} {\bibfnamefont {G.~F.}\ \bibnamefont {Chen}},
  \bibinfo {author} {\bibfnamefont {C.~E.}\ \bibnamefont {Matt}}, \bibinfo
  {author} {\bibfnamefont {F.}~\bibnamefont {Bisti}}, \bibinfo {author}
  {\bibfnamefont {V.~N.}\ \bibnamefont {Strocov}}, \bibinfo {author}
  {\bibfnamefont {J.}~\bibnamefont {Mesot}}, \bibinfo {author} {\bibfnamefont
  {Z.}~\bibnamefont {Fang}}, \bibinfo {author} {\bibfnamefont {X.}~\bibnamefont
  {Dai}}, \bibinfo {author} {\bibfnamefont {T.}~\bibnamefont {Qian}}, \bibinfo
  {author} {\bibfnamefont {M.}~\bibnamefont {Shi}}, \ and\ \bibinfo {author}
  {\bibfnamefont {H.}~\bibnamefont {Ding}},\ }\bibfield  {title} {\enquote
  {\bibinfo {title} {{Observation of Weyl nodes in TaAs}},}\ }\href
  {http://dx.doi.org/10.1038/nphys3426} {\bibfield  {journal} {\bibinfo
  {journal} {Nat Phys}\ }\textbf {\bibinfo {volume} {11}},\ \bibinfo {pages}
  {724--727} (\bibinfo {year} {2015}{\natexlab{b}})}\BibitemShut {NoStop}%
\bibitem [{\citenamefont {{Belopolski}}\ \emph {et~al.}(2015)\citenamefont
  {{Belopolski}}, \citenamefont {{Xu}}, \citenamefont {{Sanchez}},
  \citenamefont {{Chang}}, \citenamefont {{Guo}}, \citenamefont {{Neupane}},
  \citenamefont {{Zheng}}, \citenamefont {{Lee}}, \citenamefont {{Huang}},
  \citenamefont {{Bian}}, \citenamefont {{Alidoust}}, \citenamefont {{Chang}},
  \citenamefont {{Wang}}, \citenamefont {{Zhang}}, \citenamefont {{Bansil}},
  \citenamefont {{Jeng}}, \citenamefont {{Lin}}, \citenamefont {{Jia}},\ and\
  \citenamefont {{Zahid Hasan}}}]{2015arXiv150907465B}%
  \BibitemOpen
  \bibfield  {author} {\bibinfo {author} {\bibfnamefont {I.}~\bibnamefont
  {{Belopolski}}}, \bibinfo {author} {\bibfnamefont {S.-Y.}\ \bibnamefont
  {{Xu}}}, \bibinfo {author} {\bibfnamefont {D.}~\bibnamefont {{Sanchez}}},
  \bibinfo {author} {\bibfnamefont {G.}~\bibnamefont {{Chang}}}, \bibinfo
  {author} {\bibfnamefont {C.}~\bibnamefont {{Guo}}}, \bibinfo {author}
  {\bibfnamefont {M.}~\bibnamefont {{Neupane}}}, \bibinfo {author}
  {\bibfnamefont {H.}~\bibnamefont {{Zheng}}}, \bibinfo {author} {\bibfnamefont
  {C.-C.}\ \bibnamefont {{Lee}}}, \bibinfo {author} {\bibfnamefont {S.-M.}\
  \bibnamefont {{Huang}}}, \bibinfo {author} {\bibfnamefont {G.}~\bibnamefont
  {{Bian}}}, \bibinfo {author} {\bibfnamefont {N.}~\bibnamefont {{Alidoust}}},
  \bibinfo {author} {\bibfnamefont {T.-R.}\ \bibnamefont {{Chang}}}, \bibinfo
  {author} {\bibfnamefont {B.}~\bibnamefont {{Wang}}}, \bibinfo {author}
  {\bibfnamefont {X.}~\bibnamefont {{Zhang}}}, \bibinfo {author} {\bibfnamefont
  {A.}~\bibnamefont {{Bansil}}}, \bibinfo {author} {\bibfnamefont {H.-T.}\
  \bibnamefont {{Jeng}}}, \bibinfo {author} {\bibfnamefont {H.}~\bibnamefont
  {{Lin}}}, \bibinfo {author} {\bibfnamefont {S.}~\bibnamefont {{Jia}}}, \ and\
  \bibinfo {author} {\bibfnamefont {M.}~\bibnamefont {{Zahid Hasan}}},\
  }\bibfield  {title} {\enquote {\bibinfo {title} {{Observation of surface
  states derived from topological Fermi arcs in the Weyl semimetal NbP}},}\
  }\href@noop {} {\bibfield  {journal} {\bibinfo  {journal} {ArXiv e-prints}\ }
  (\bibinfo {year} {2015})},\ \Eprint {http://arxiv.org/abs/1509.07465}
  {arXiv:1509.07465 [cond-mat.mes-hall]} \BibitemShut {NoStop}%
\bibitem [{\citenamefont {Xu}\ \emph {et~al.}(2015)\citenamefont {Xu},
  \citenamefont {Alidoust}, \citenamefont {Belopolski}, \citenamefont {Yuan},
  \citenamefont {Bian}, \citenamefont {Chang}, \citenamefont {Zheng},
  \citenamefont {Strocov}, \citenamefont {Sanchez}, \citenamefont {Chang},
  \citenamefont {Zhang}, \citenamefont {Mou}, \citenamefont {Wu}, \citenamefont
  {Huang}, \citenamefont {Lee}, \citenamefont {Huang}, \citenamefont {Wang},
  \citenamefont {Bansil}, \citenamefont {Jeng}, \citenamefont {Neupert},
  \citenamefont {Kaminski}, \citenamefont {Lin}, \citenamefont {Jia},\ and\
  \citenamefont {Zahid~Hasan}}]{Xu:2015aa}%
  \BibitemOpen
  \bibfield  {author} {\bibinfo {author} {\bibfnamefont {Su-Yang}\ \bibnamefont
  {Xu}}, \bibinfo {author} {\bibfnamefont {Nasser}\ \bibnamefont {Alidoust}},
  \bibinfo {author} {\bibfnamefont {Ilya}\ \bibnamefont {Belopolski}}, \bibinfo
  {author} {\bibfnamefont {Zhujun}\ \bibnamefont {Yuan}}, \bibinfo {author}
  {\bibfnamefont {Guang}\ \bibnamefont {Bian}}, \bibinfo {author}
  {\bibfnamefont {Tay-Rong}\ \bibnamefont {Chang}}, \bibinfo {author}
  {\bibfnamefont {Hao}\ \bibnamefont {Zheng}}, \bibinfo {author} {\bibfnamefont
  {Vladimir~N.}\ \bibnamefont {Strocov}}, \bibinfo {author} {\bibfnamefont
  {Daniel~S.}\ \bibnamefont {Sanchez}}, \bibinfo {author} {\bibfnamefont
  {Guoqing}\ \bibnamefont {Chang}}, \bibinfo {author} {\bibfnamefont
  {Chenglong}\ \bibnamefont {Zhang}}, \bibinfo {author} {\bibfnamefont
  {Daixiang}\ \bibnamefont {Mou}}, \bibinfo {author} {\bibfnamefont {Yun}\
  \bibnamefont {Wu}}, \bibinfo {author} {\bibfnamefont {Lunan}\ \bibnamefont
  {Huang}}, \bibinfo {author} {\bibfnamefont {Chi-Cheng}\ \bibnamefont {Lee}},
  \bibinfo {author} {\bibfnamefont {Shin-Ming}\ \bibnamefont {Huang}}, \bibinfo
  {author} {\bibfnamefont {BaoKai}\ \bibnamefont {Wang}}, \bibinfo {author}
  {\bibfnamefont {Arun}\ \bibnamefont {Bansil}}, \bibinfo {author}
  {\bibfnamefont {Horng-Tay}\ \bibnamefont {Jeng}}, \bibinfo {author}
  {\bibfnamefont {Titus}\ \bibnamefont {Neupert}}, \bibinfo {author}
  {\bibfnamefont {Adam}\ \bibnamefont {Kaminski}}, \bibinfo {author}
  {\bibfnamefont {Hsin}\ \bibnamefont {Lin}}, \bibinfo {author} {\bibfnamefont
  {Shuang}\ \bibnamefont {Jia}}, \ and\ \bibinfo {author} {\bibfnamefont
  {M.}~\bibnamefont {Zahid~Hasan}},\ }\bibfield  {title} {\enquote {\bibinfo
  {title} {{Discovery of a Weyl fermion state with Fermi arcs in niobium
  arsenide}},}\ }\href {http://dx.doi.org/10.1038/nphys3437} {\bibfield
  {journal} {\bibinfo  {journal} {Nat Phys}\ }\textbf {\bibinfo {volume}
  {11}},\ \bibinfo {pages} {748--754} (\bibinfo {year} {2015})}\BibitemShut
  {NoStop}%
\bibitem [{\citenamefont {Fukushima}\ \emph {et~al.}(2008)\citenamefont
  {Fukushima}, \citenamefont {Kharzeev},\ and\ \citenamefont
  {Warringa}}]{PhysRevD.78.074033}%
  \BibitemOpen
  \bibfield  {author} {\bibinfo {author} {\bibfnamefont {Kenji}\ \bibnamefont
  {Fukushima}}, \bibinfo {author} {\bibfnamefont {Dmitri~E.}\ \bibnamefont
  {Kharzeev}}, \ and\ \bibinfo {author} {\bibfnamefont {Harmen~J.}\
  \bibnamefont {Warringa}},\ }\bibfield  {title} {\enquote {\bibinfo {title}
  {{Chiral magnetic effect}},}\ }\href {\doibase 10.1103/PhysRevD.78.074033}
  {\bibfield  {journal} {\bibinfo  {journal} {Phys. Rev. D}\ }\textbf {\bibinfo
  {volume} {78}},\ \bibinfo {pages} {074033} (\bibinfo {year}
  {2008})}\BibitemShut {NoStop}%
\bibitem [{\citenamefont {Burkov}(2015)}]{0953-8984-27-11-113201}%
  \BibitemOpen
  \bibfield  {author} {\bibinfo {author} {\bibfnamefont {A~A}\ \bibnamefont
  {Burkov}},\ }\bibfield  {title} {\enquote {\bibinfo {title} {{Chiral anomaly
  and transport in Weyl metals}},}\ }\href
  {http://stacks.iop.org/0953-8984/27/i=11/a=113201} {\bibfield  {journal}
  {\bibinfo  {journal} {Journal of Physics: Condensed Matter}\ }\textbf
  {\bibinfo {volume} {27}},\ \bibinfo {pages} {113201} (\bibinfo {year}
  {2015})}\BibitemShut {NoStop}%
\bibitem [{\citenamefont {Grushin}(2012)}]{PhysRevD.86.045001}%
  \BibitemOpen
  \bibfield  {author} {\bibinfo {author} {\bibfnamefont {Adolfo~G.}\
  \bibnamefont {Grushin}},\ }\bibfield  {title} {\enquote {\bibinfo {title}
  {{Consequences of a condensed matter realization of Lorentz-violating QED in
  Weyl semi-metals}},}\ }\href {\doibase 10.1103/PhysRevD.86.045001} {\bibfield
   {journal} {\bibinfo  {journal} {Phys. Rev. D}\ }\textbf {\bibinfo {volume}
  {86}},\ \bibinfo {pages} {045001} (\bibinfo {year} {2012})}\BibitemShut
  {NoStop}%
\bibitem [{\citenamefont {Landsteiner}(2014)}]{PhysRevB.89.075124}%
  \BibitemOpen
  \bibfield  {author} {\bibinfo {author} {\bibfnamefont {Karl}\ \bibnamefont
  {Landsteiner}},\ }\bibfield  {title} {\enquote {\bibinfo {title} {{Anomalous
  transport of Weyl fermions in Weyl semimetals}},}\ }\href {\doibase
  10.1103/PhysRevB.89.075124} {\bibfield  {journal} {\bibinfo  {journal} {Phys.
  Rev. B}\ }\textbf {\bibinfo {volume} {89}},\ \bibinfo {pages} {075124}
  (\bibinfo {year} {2014})}\BibitemShut {NoStop}%
\bibitem [{\citenamefont {Chernodub}\ \emph {et~al.}(2014)\citenamefont
  {Chernodub}, \citenamefont {Cortijo}, \citenamefont {Grushin}, \citenamefont
  {Landsteiner},\ and\ \citenamefont {Vozmediano}}]{PhysRevB.89.081407}%
  \BibitemOpen
  \bibfield  {author} {\bibinfo {author} {\bibfnamefont {Maxim~N.}\
  \bibnamefont {Chernodub}}, \bibinfo {author} {\bibfnamefont {Alberto}\
  \bibnamefont {Cortijo}}, \bibinfo {author} {\bibfnamefont {Adolfo~G.}\
  \bibnamefont {Grushin}}, \bibinfo {author} {\bibfnamefont {Karl}\
  \bibnamefont {Landsteiner}}, \ and\ \bibinfo {author} {\bibfnamefont
  {Mar\'{\i}a A.~H.}\ \bibnamefont {Vozmediano}},\ }\bibfield  {title}
  {\enquote {\bibinfo {title} {{Condensed matter realization of the axial
  magnetic effect}},}\ }\href {\doibase 10.1103/PhysRevB.89.081407} {\bibfield
  {journal} {\bibinfo  {journal} {Phys. Rev. B}\ }\textbf {\bibinfo {volume}
  {89}},\ \bibinfo {pages} {081407} (\bibinfo {year} {2014})}\BibitemShut
  {NoStop}%
\bibitem [{\citenamefont {Goswami}\ and\ \citenamefont
  {Tewari}(2013)}]{Pallab_noncentro}%
  \BibitemOpen
  \bibfield  {author} {\bibinfo {author} {\bibfnamefont {Pallab}\ \bibnamefont
  {Goswami}}\ and\ \bibinfo {author} {\bibfnamefont {Sumanta}\ \bibnamefont
  {Tewari}},\ }\bibfield  {title} {\enquote {\bibinfo {title} {{Chiral magnetic
  effect of Weyl fermions and its applications to cubic noncentrosymmetric
  metals}},}\ }\href@noop {} {\bibfield  {journal} {\bibinfo  {journal} {ArXiv
  e-prints}\ } (\bibinfo {year} {2013})},\ \Eprint
  {http://arxiv.org/abs/1311.1506} {arXiv:1311.1506 [cond-mat.mes-hall]}
  \BibitemShut {NoStop}%
%%CITATION = ARXIV:1311.1506;%%
\bibitem [{\citenamefont {Goswami}\ \emph
  {et~al.}(2015{\natexlab{a}})\citenamefont {Goswami}, \citenamefont {Sharma},\
  and\ \citenamefont {Tewari}}]{PhysRevB.92.161110}%
  \BibitemOpen
  \bibfield  {author} {\bibinfo {author} {\bibfnamefont {Pallab}\ \bibnamefont
  {Goswami}}, \bibinfo {author} {\bibfnamefont {Girish}\ \bibnamefont
  {Sharma}}, \ and\ \bibinfo {author} {\bibfnamefont {Sumanta}\ \bibnamefont
  {Tewari}},\ }\bibfield  {title} {\enquote {\bibinfo {title} {{Optical
  activity as a test for dynamic chiral magnetic effect of Weyl semimetals}},}\
  }\href {\doibase 10.1103/PhysRevB.92.161110} {\bibfield  {journal} {\bibinfo
  {journal} {Phys. Rev. B}\ }\textbf {\bibinfo {volume} {92}},\ \bibinfo
  {pages} {161110} (\bibinfo {year} {2015}{\natexlab{a}})}\BibitemShut
  {NoStop}%
\bibitem [{\citenamefont {Chang}\ and\ \citenamefont
  {Yang}(2015)}]{PhysRevB.91.115203}%
  \BibitemOpen
  \bibfield  {author} {\bibinfo {author} {\bibfnamefont {Ming-Che}\
  \bibnamefont {Chang}}\ and\ \bibinfo {author} {\bibfnamefont {Min-Fong}\
  \bibnamefont {Yang}},\ }\bibfield  {title} {\enquote {\bibinfo {title}
  {{Chiral magnetic effect in a two-band lattice model of Weyl semimetal}},}\
  }\href {\doibase 10.1103/PhysRevB.91.115203} {\bibfield  {journal} {\bibinfo
  {journal} {Phys. Rev. B}\ }\textbf {\bibinfo {volume} {91}},\ \bibinfo
  {pages} {115203} (\bibinfo {year} {2015})}\BibitemShut {NoStop}%
\bibitem [{\citenamefont {Kharzeev}(2014)}]{Kharzeev2014133}%
  \BibitemOpen
  \bibfield  {author} {\bibinfo {author} {\bibfnamefont {Dmitri~E.}\
  \bibnamefont {Kharzeev}},\ }\bibfield  {title} {\enquote {\bibinfo {title}
  {{The Chiral Magnetic Effect and anomaly-induced transport}},}\ }\href
  {\doibase http://dx.doi.org/10.1016/j.ppnp.2014.01.002} {\bibfield  {journal}
  {\bibinfo  {journal} {Progress in Particle and Nuclear Physics}\ }\textbf
  {\bibinfo {volume} {75}},\ \bibinfo {pages} {133 -- 151} (\bibinfo {year}
  {2014})}\BibitemShut {NoStop}%
\bibitem [{\citenamefont {Goswami}\ \emph
  {et~al.}(2015{\natexlab{b}})\citenamefont {Goswami}, \citenamefont {Pixley},\
  and\ \citenamefont {Das~Sarma}}]{PhysRevB.92.075205}%
  \BibitemOpen
  \bibfield  {author} {\bibinfo {author} {\bibfnamefont {Pallab}\ \bibnamefont
  {Goswami}}, \bibinfo {author} {\bibfnamefont {J.~H.}\ \bibnamefont {Pixley}},
  \ and\ \bibinfo {author} {\bibfnamefont {S.}~\bibnamefont {Das~Sarma}},\
  }\bibfield  {title} {\enquote {\bibinfo {title} {{Axial anomaly and
  longitudinal magnetoresistance of a generic three-dimensional metal}},}\
  }\href {\doibase 10.1103/PhysRevB.92.075205} {\bibfield  {journal} {\bibinfo
  {journal} {Phys. Rev. B}\ }\textbf {\bibinfo {volume} {92}},\ \bibinfo
  {pages} {075205} (\bibinfo {year} {2015}{\natexlab{b}})}\BibitemShut
  {NoStop}%
\bibitem [{\citenamefont {{Kikugawa}}\ \emph {et~al.}(2014)\citenamefont
  {{Kikugawa}}, \citenamefont {{Goswami}}, \citenamefont {{Kiswandhi}},
  \citenamefont {{Choi}}, \citenamefont {{Graf}}, \citenamefont {{Baumbach}},
  \citenamefont {{Brooks}}, \citenamefont {{Sugii}}, \citenamefont {{Iida}},
  \citenamefont {{Nishio}}, \citenamefont {{Uji}}, \citenamefont {{Terashima}},
  \citenamefont {{Rourke}}, \citenamefont {{Hussey}}, \citenamefont
  {{Takatsu}}, \citenamefont {{Yonezawa}}, \citenamefont {{Maeno}},\ and\
  \citenamefont {{Balicas}}}]{2014arXiv1412.5168K}%
  \BibitemOpen
  \bibfield  {author} {\bibinfo {author} {\bibfnamefont {N.}~\bibnamefont
  {{Kikugawa}}}, \bibinfo {author} {\bibfnamefont {P.}~\bibnamefont
  {{Goswami}}}, \bibinfo {author} {\bibfnamefont {A.}~\bibnamefont
  {{Kiswandhi}}}, \bibinfo {author} {\bibfnamefont {E.~S.}\ \bibnamefont
  {{Choi}}}, \bibinfo {author} {\bibfnamefont {D.}~\bibnamefont {{Graf}}},
  \bibinfo {author} {\bibfnamefont {R.~E.}\ \bibnamefont {{Baumbach}}},
  \bibinfo {author} {\bibfnamefont {J.~S.}\ \bibnamefont {{Brooks}}}, \bibinfo
  {author} {\bibfnamefont {K.}~\bibnamefont {{Sugii}}}, \bibinfo {author}
  {\bibfnamefont {Y.}~\bibnamefont {{Iida}}}, \bibinfo {author} {\bibfnamefont
  {M.}~\bibnamefont {{Nishio}}}, \bibinfo {author} {\bibfnamefont
  {S.}~\bibnamefont {{Uji}}}, \bibinfo {author} {\bibfnamefont
  {T.}~\bibnamefont {{Terashima}}}, \bibinfo {author} {\bibfnamefont
  {P.~M.~C.}\ \bibnamefont {{Rourke}}}, \bibinfo {author} {\bibfnamefont
  {N.~E.}\ \bibnamefont {{Hussey}}}, \bibinfo {author} {\bibfnamefont
  {H.}~\bibnamefont {{Takatsu}}}, \bibinfo {author} {\bibfnamefont
  {S.}~\bibnamefont {{Yonezawa}}}, \bibinfo {author} {\bibfnamefont
  {Y.}~\bibnamefont {{Maeno}}}, \ and\ \bibinfo {author} {\bibfnamefont
  {L.}~\bibnamefont {{Balicas}}},\ }\bibfield  {title} {\enquote {\bibinfo
  {title} {{Inter-planar coupling dependent magnetoresistivity in high purity
  layered metals}},}\ }\href@noop {} {\bibfield  {journal} {\bibinfo  {journal}
  {ArXiv e-prints}\ } (\bibinfo {year} {2014})},\ \Eprint
  {http://arxiv.org/abs/1412.5168} {arXiv:1412.5168 [cond-mat.mes-hall]}
  \BibitemShut {NoStop}%
\bibitem [{\citenamefont {Xiong}\ \emph {et~al.}(2015)\citenamefont {Xiong},
  \citenamefont {Kushwaha}, \citenamefont {Liang}, \citenamefont {Krizan},
  \citenamefont {Hirschberger}, \citenamefont {Wang}, \citenamefont {Cava},\
  and\ \citenamefont {Ong}}]{Xiong23102015}%
  \BibitemOpen
  \bibfield  {author} {\bibinfo {author} {\bibfnamefont {Jun}\ \bibnamefont
  {Xiong}}, \bibinfo {author} {\bibfnamefont {Satya~K.}\ \bibnamefont
  {Kushwaha}}, \bibinfo {author} {\bibfnamefont {Tian}\ \bibnamefont {Liang}},
  \bibinfo {author} {\bibfnamefont {Jason~W.}\ \bibnamefont {Krizan}}, \bibinfo
  {author} {\bibfnamefont {Max}\ \bibnamefont {Hirschberger}}, \bibinfo
  {author} {\bibfnamefont {Wudi}\ \bibnamefont {Wang}}, \bibinfo {author}
  {\bibfnamefont {R.~J.}\ \bibnamefont {Cava}}, \ and\ \bibinfo {author}
  {\bibfnamefont {N.~P.}\ \bibnamefont {Ong}},\ }\bibfield  {title} {\enquote
  {\bibinfo {title} {{Evidence for the chiral anomaly in the Dirac semimetal
  Na3Bi}},}\ }\href {\doibase 10.1126/science.aac6089} {\bibfield  {journal}
  {\bibinfo  {journal} {Science}\ }\textbf {\bibinfo {volume} {350}},\ \bibinfo
  {pages} {413--416} (\bibinfo {year} {2015})}\BibitemShut {NoStop}%
\bibitem [{\citenamefont {Huang}\ \emph {et~al.}(2015)\citenamefont {Huang},
  \citenamefont {Zhao}, \citenamefont {Long}, \citenamefont {Wang},
  \citenamefont {Chen}, \citenamefont {Yang}, \citenamefont {Liang},
  \citenamefont {Xue}, \citenamefont {Weng}, \citenamefont {Fang},
  \citenamefont {Dai},\ and\ \citenamefont {Chen}}]{PhysRevX.5.031023}%
  \BibitemOpen
  \bibfield  {author} {\bibinfo {author} {\bibfnamefont {Xiaochun}\
  \bibnamefont {Huang}}, \bibinfo {author} {\bibfnamefont {Lingxiao}\
  \bibnamefont {Zhao}}, \bibinfo {author} {\bibfnamefont {Yujia}\ \bibnamefont
  {Long}}, \bibinfo {author} {\bibfnamefont {Peipei}\ \bibnamefont {Wang}},
  \bibinfo {author} {\bibfnamefont {Dong}\ \bibnamefont {Chen}}, \bibinfo
  {author} {\bibfnamefont {Zhanhai}\ \bibnamefont {Yang}}, \bibinfo {author}
  {\bibfnamefont {Hui}\ \bibnamefont {Liang}}, \bibinfo {author} {\bibfnamefont
  {Mianqi}\ \bibnamefont {Xue}}, \bibinfo {author} {\bibfnamefont {Hongming}\
  \bibnamefont {Weng}}, \bibinfo {author} {\bibfnamefont {Zhong}\ \bibnamefont
  {Fang}}, \bibinfo {author} {\bibfnamefont {Xi}~\bibnamefont {Dai}}, \ and\
  \bibinfo {author} {\bibfnamefont {Genfu}\ \bibnamefont {Chen}},\ }\bibfield
  {title} {\enquote {\bibinfo {title} {{Observation of the
  Chiral-Anomaly-Induced Negative Magnetoresistance in 3D Weyl Semimetal
  TaAs}},}\ }\href {\doibase 10.1103/PhysRevX.5.031023} {\bibfield  {journal}
  {\bibinfo  {journal} {Phys. Rev. X}\ }\textbf {\bibinfo {volume} {5}},\
  \bibinfo {pages} {031023} (\bibinfo {year} {2015})}\BibitemShut {NoStop}%
\bibitem [{\citenamefont {Son}\ and\ \citenamefont
  {Spivak}(2013)}]{PhysRevB.88.104412}%
  \BibitemOpen
  \bibfield  {author} {\bibinfo {author} {\bibfnamefont {D.~T.}\ \bibnamefont
  {Son}}\ and\ \bibinfo {author} {\bibfnamefont {B.~Z.}\ \bibnamefont
  {Spivak}},\ }\bibfield  {title} {\enquote {\bibinfo {title} {{Chiral anomaly
  and classical negative magnetoresistance of Weyl metals}},}\ }\href {\doibase
  10.1103/PhysRevB.88.104412} {\bibfield  {journal} {\bibinfo  {journal} {Phys.
  Rev. B}\ }\textbf {\bibinfo {volume} {88}},\ \bibinfo {pages} {104412}
  (\bibinfo {year} {2013})}\BibitemShut {NoStop}%
\bibitem [{\citenamefont {Vazifeh}\ and\ \citenamefont
  {Franz}(2013)}]{PhysRevLett.111.027201}%
  \BibitemOpen
  \bibfield  {author} {\bibinfo {author} {\bibfnamefont {M.~M.}\ \bibnamefont
  {Vazifeh}}\ and\ \bibinfo {author} {\bibfnamefont {M.}~\bibnamefont
  {Franz}},\ }\bibfield  {title} {\enquote {\bibinfo {title} {{Electromagnetic
  Response of Weyl Semimetals}},}\ }\href {\doibase
  10.1103/PhysRevLett.111.027201} {\bibfield  {journal} {\bibinfo  {journal}
  {Phys. Rev. Lett.}\ }\textbf {\bibinfo {volume} {111}},\ \bibinfo {pages}
  {027201} (\bibinfo {year} {2013})}\BibitemShut {NoStop}%
\bibitem [{\citenamefont {Chen}\ \emph {et~al.}(2013)\citenamefont {Chen},
  \citenamefont {Wu},\ and\ \citenamefont {Burkov}}]{PhysRevB.88.125105}%
  \BibitemOpen
  \bibfield  {author} {\bibinfo {author} {\bibfnamefont {Y.}~\bibnamefont
  {Chen}}, \bibinfo {author} {\bibfnamefont {Si}~\bibnamefont {Wu}}, \ and\
  \bibinfo {author} {\bibfnamefont {A.~A.}\ \bibnamefont {Burkov}},\ }\bibfield
   {title} {\enquote {\bibinfo {title} {{Axion response in Weyl semimetals}},}\
  }\href {\doibase 10.1103/PhysRevB.88.125105} {\bibfield  {journal} {\bibinfo
  {journal} {Phys. Rev. B}\ }\textbf {\bibinfo {volume} {88}},\ \bibinfo
  {pages} {125105} (\bibinfo {year} {2013})}\BibitemShut {NoStop}%
\bibitem [{\citenamefont {Wilczek}(1987)}]{PhysRevLett.58.1799}%
  \BibitemOpen
  \bibfield  {author} {\bibinfo {author} {\bibfnamefont {Frank}\ \bibnamefont
  {Wilczek}},\ }\bibfield  {title} {\enquote {\bibinfo {title} {{Two
  applications of axion electrodynamics}},}\ }\href {\doibase
  10.1103/PhysRevLett.58.1799} {\bibfield  {journal} {\bibinfo  {journal}
  {Phys. Rev. Lett.}\ }\textbf {\bibinfo {volume} {58}},\ \bibinfo {pages}
  {1799--1802} (\bibinfo {year} {1987})}\BibitemShut {NoStop}%
\bibitem [{\citenamefont {Ooguri}\ and\ \citenamefont
  {Oshikawa}(2012)}]{PhysRevLett.108.161803}%
  \BibitemOpen
  \bibfield  {author} {\bibinfo {author} {\bibfnamefont {Hirosi}\ \bibnamefont
  {Ooguri}}\ and\ \bibinfo {author} {\bibfnamefont {Masaki}\ \bibnamefont
  {Oshikawa}},\ }\bibfield  {title} {\enquote {\bibinfo {title} {{Instability
  in Magnetic Materials with a Dynamical Axion Field}},}\ }\href {\doibase
  10.1103/PhysRevLett.108.161803} {\bibfield  {journal} {\bibinfo  {journal}
  {Phys. Rev. Lett.}\ }\textbf {\bibinfo {volume} {108}},\ \bibinfo {pages}
  {161803} (\bibinfo {year} {2012})}\BibitemShut {NoStop}%
\bibitem [{\citenamefont {Kharzeev}\ and\ \citenamefont
  {Yee}(2013)}]{PhysRevB.88.115119}%
  \BibitemOpen
  \bibfield  {author} {\bibinfo {author} {\bibfnamefont {Dmitri~E.}\
  \bibnamefont {Kharzeev}}\ and\ \bibinfo {author} {\bibfnamefont {Ho-Ung}\
  \bibnamefont {Yee}},\ }\bibfield  {title} {\enquote {\bibinfo {title}
  {{Anomaly induced chiral magnetic current in a Weyl semimetal: Chiral
  electronics}},}\ }\href {\doibase 10.1103/PhysRevB.88.115119} {\bibfield
  {journal} {\bibinfo  {journal} {Phys. Rev. B}\ }\textbf {\bibinfo {volume}
  {88}},\ \bibinfo {pages} {115119} (\bibinfo {year} {2013})}\BibitemShut
  {NoStop}%
\end{thebibliography}%
\end{document}